\documentclass[5p,times,twocolumn,final]{elsarticle} 

\usepackage{graphicx}
\usepackage{amsmath,amssymb}

\DeclareMathOperator{\im}{Im}

\DeclareMathOperator{\sech}{sech}
\DeclareMathOperator{\cn}{cn}
\DeclareMathOperator{\Ai}{Ai}
\newcommand{\ud}{\mathrm{d}}

\journal{Physics Letters A}

\begin{document}

\begin{frontmatter}

\title{Interactions and Asymptotics of Dispersive Shock Waves ---  Korteweg--de~Vries Equation}

\author{Mark J.\ Ablowitz}
\author{Douglas E.\ Baldwin}
\ead{dsw@douglasbaldwin.com}
\ead[url]{http://www.douglasbaldwin.com}
\address{Department of Applied Mathematics, University of Colorado, Boulder, Colorado, 80309-0526, USA}

\begin{abstract}
The long-time asymptotic solution of the Korteweg--deVries equation for general, step-like initial data is analyzed. 
Each sub-step in well-separated, multi-step data forms its own single dispersive shock wave (DSW); 
at intermediate times these DSWs interact and develop multiphase dynamics. 
Using the inverse scattering transform and matched-asymptotic analysis it is shown that the DSWs merge to form a single-phase DSW, which is the `largest' one possible for the boundary data. 
This is similar to interacting viscous shock waves (VSW) that are modeled with Burgers' equation, where only the single, largest-possible VSW remains after a long time.
\end{abstract}


\end{frontmatter}

\section{Introduction}

Dispersive shock waves (DSWs) appear when dispersion dominates dissipation for step-like data; 
they have been seen in plasmas \cite{Taylor1970}, fluids (e.g., undular bores) \cite{Smyth1988,LighthillBook}, superfluids \cite{Dutton2001,Simula2005,Hoefer2006,Chang2008}, and optics \cite{Wan2007,Jia2007,Ghofraniha2007,Conti2009}. 
The Korteweg--de~Vries (KdV) equation is the leading-order asymptotic equation for weakly dispersive and weakly nonlinear systems \cite{Ablowitzbook2011}. 
Each step in well-separated, multi-step data forms its own DSW; 
these DSWs interact and develop multiphase dynamics \cite{ABH2009}. 
Here we show that these DSWs merge in the long-time limit to form a single-phase DSW;  
the boundary data determine its form and the initial data determine its location. 
We find this long-time asymptotic solution using the inverse scattering transform (IST) and matched-asymptotic analysis. 
This merging of shock waves is similar to viscous shock waves (VSWs), where dissipation dominates dispersion and only a single VSW remains in the long-time limit. 
While the linear KdV equation's solution for step-like data in the middle region has width $\mathcal{O}(t^{1/3})$ (see \S\ref{sec:linearKdV}) 
and the KdV equation's solution for vanishing data has a collisionless shock of width $\mathcal{O}[t^{1/3}(\log t)^{2/3}]$ (see \S\ref{sec:vanishing}), 
the DSW that we find has width $\mathcal{O}(t)$ (\S\ref{sec:DSW}). 
We anticipate that our IST and matched-asymptotic procedure will be applied to other integrable nonlinear partial differential equations (PDEs) with step-like data.

Here we consider DSWs described by the KdV equation --- the leading-order asymptotic equation for systems with weak dispersion and weak, quadratic nonlinearity.   
The KdV equation, in dimensionless form, is 
\begin{equation} \label{eq:KdV}
  u_t + uu_x + \varepsilon^2u_{xxx} = 0, 
\end{equation} 
where subscripts denote partial derivatives. 
We require that $u = u(x,t)$ goes rapidly to the boundary conditions 
\begin{equation} \label{eq:uBC} 
	\lim_{x\to-\infty} u = 0 \quad\text{and}\quad
	\lim_{x\to+\infty} u = - 6c^2,
\end{equation}
where $\varepsilon$ and $c$ are real, positive constants. 
Here, $\varepsilon$ corresponds to the size of the regularizing dispersive effects.
(Since the KdV equation is Galilean invariant, we can transform any boundary conditions where $\lim_{x\to-\infty} u > \lim_{x\to+\infty} u$ to these boundary conditions.) 
We use the IST method (see \cite{AKNS1974,DeiftTrubowitz1979,AblowitzSegurBook,Ablowitz1991}) and matched-asymptotic expansions (see \cite{AblowitzSegur1977,Segur1981}) to find a large-time asymptotic solution.

\begin{figure}
	\begin{tabular}{cc}
		\includegraphics[width=0.45\hsize]{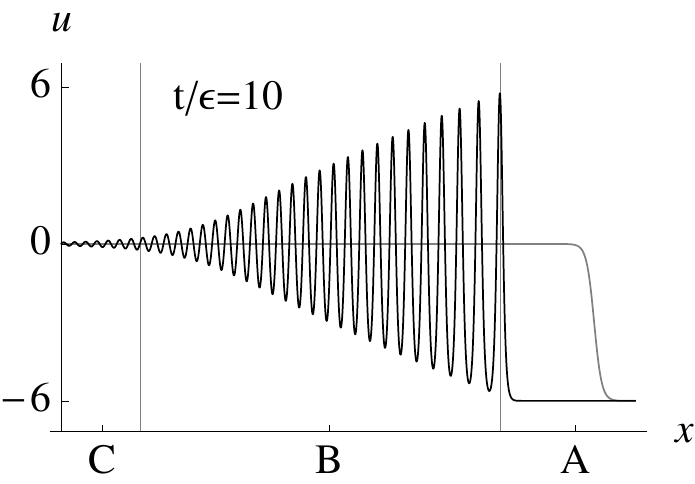} & 
		\includegraphics[width=0.45\hsize]{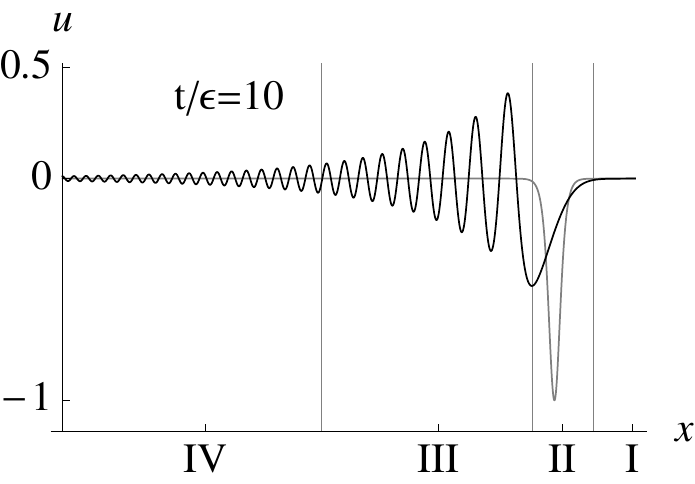} \\
		(a) & (b) \\
	\end{tabular}
	\caption{\label{fig:decay} 
	Numerical solutions of the KdV equation for the step and the vanishing initial data shown in gray. 
	(a)~For step data, there are three basic regions: 
		exponential decay in region~A, 
		the DSW in region~B with width $\mathcal{O}(t)$ and height $\mathcal{O}(1)$, and
		an oscillating tail in region~C.
	(b)~For vanishing data, there are four basic regions (see \cite{AblowitzSegur1977}). 
		The collisonless shock in region~III, which is analogous to the DSW in region~B, has width $\mathcal{O}[t^{1/3}(\log t)^{2/3}]$ and height $\mathcal{O}[(\log t)^{1/2}t^{-2/3}]$. }
\end{figure}

\begin{figure}
	\begin{tabular}{cc}
		\includegraphics[width=0.45\linewidth]{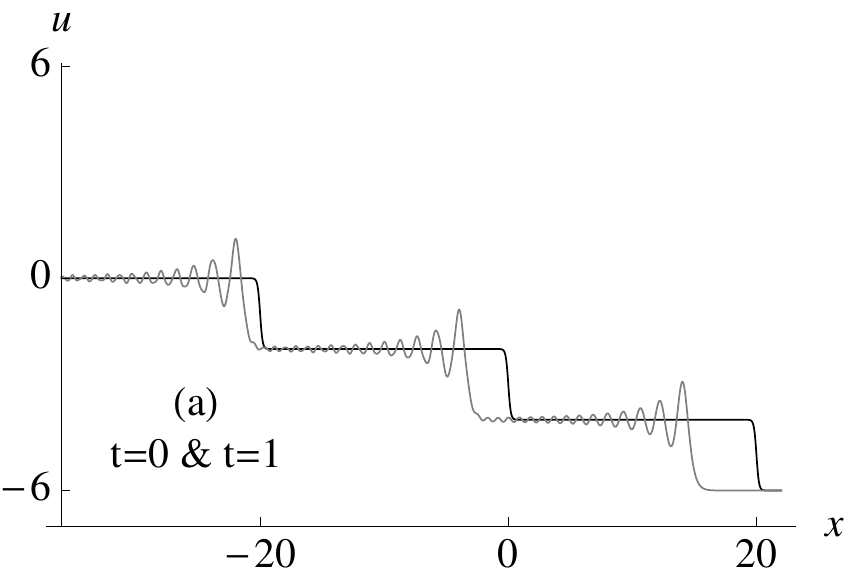} & 
		\includegraphics[width=0.45\linewidth]{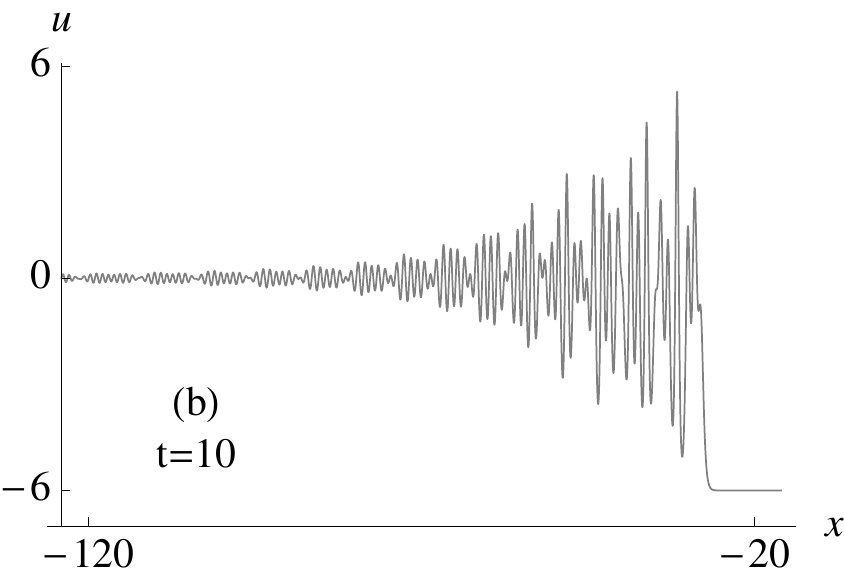} \\
		\includegraphics[width=0.45\linewidth]{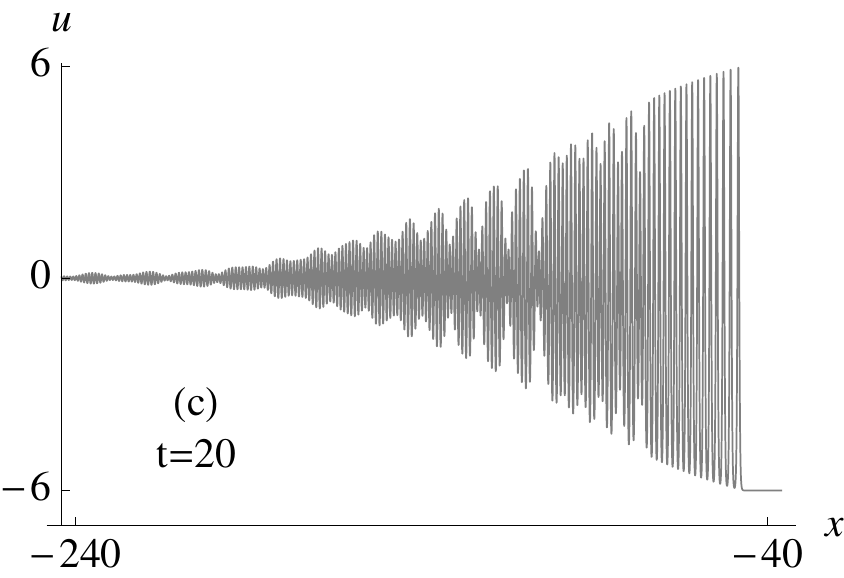} & 
		\includegraphics[width=0.45\linewidth]{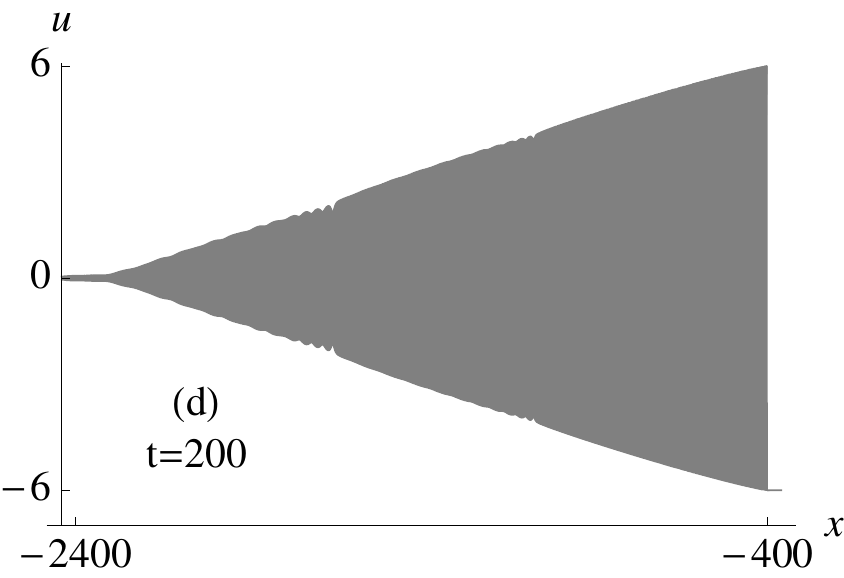} \\
	\end{tabular}
	\caption{\label{fig:numerics} A numerical simulation (using the scheme in \cite{ABH2009}) of three well-separated steps (at $t=0$) that:
	(a)~form three single-phase DSWs at $t=1$;
	(b) and (c) interact strongly and exhibit multiphase dynamics;
	and (d)~eventually merge to form a single-phase DSW.
	Here, $\varepsilon^2 = 0.1$ and $c = 1$.}
\end{figure}

Single-step data, such as a Heaviside function, evolve to form a single DSW. 
This DSW has three basic regions (Fig.~\ref{fig:decay}a): 
a rapidly decaying region to the right of the DSW (region~A); 
the central DSW (region~B), which is a slowly varying cnoidal wave with a soliton train on its right and an oscillatory tail on its left; 
and a decaying, oscillatory region to the left of the DSW (region~C). 
Similarly, each step in well-separated, multi-step data forms its own DSW (Fig.~\ref{fig:numerics}a); 
these DSWs eventually interact strongly and at intermediate times exhibit multiphase dynamics (Figs. \ref{fig:numerics}b and \ref{fig:numerics}c). 
In this letter we show, in the long-time limit, that these DSWs eventually merge to form a single-phase DSW (Fig.~\ref{fig:numerics}d). 

We use IST and matched-asymptotic methods to show that general, step-like data go to a single-phase DSW in the large-time limit. 
Individual, single-phase DSWs have been extensively studied using wave averaging techniques (see \cite{Gurevich1974,Kamchatnov2000,El2002,El2005c}), often referred to as Whitham theory \cite{Whitham1965,Whitham1974}.  
The evolution of two-phase DSWs to a single-phase DSW was investigated by \citet{Grava2002} in the zero-dispersion ($\varepsilon \to 0$) finite-time limit using Whitham theory and by \citet{ABH2009} in the fixed-dispersion long-time limit using numerical and asymptotic methods. 
Both zero-dispersion and large-time are important, but different, limits; 
here we study the large-time limit with fixed dispersion. 
By using the IST method we find the asymptotic solution directly; 
in Whitham theory, we must evolve the solution through intermediate times.
Therefore, we can investigate general, step-like initial data and the interaction of DSWs without having to find the solution at intermediate times. 

The IST theory for step-like initial conditions was studied in \cite{Hruslov1976} and \cite{Cohen1984,Cohen1985}. 
\citet{Hruslov1976}, based on \cite{BuslaevFomin}, gives the Gel'fand--Lev\-i\-tan--Mar\-chen\-ko (GLM) integral equations and investigates the soliton train associated with the DSW.
\citet{Cohen1984} and \citet{Cohen1985}, using the methods in \cite{DeiftTrubowitz1979,BuslaevFomin}, 
rigorously studied the properties of the scattering data, 
rederived the GLM integral equations, 
and analyzed existence of solutions corresponding to certain initial conditions. 
We state the IST results that we need to find our asymptotic solution in \S\ref{sec:IST}.

From these IST results, 
we use and suitably modify the methods in \cite{AblowitzSegur1977,Segur1981} to find our long-time asymptotic solution. 
Ablowitz and Segur \cite{AblowitzSegur1977,Segur1981} developed the IST and matched-asymptotic methods for vanishing data (where $c = 0$);
we modify them for step-like data (where $c\neq 0$).
The results of the long-time asymptotic analysis in this letter are new. 
There are elegant and powerful asymptotic methods based on Riemann--Hilbert problems that depend on a parameter (here time, $t$); 
they have been used to find the long-time asymptotic solution for vanishing data (see \cite{DeiftZhou,Deift1997}) --- 
see also \cite{Buckingham2007} for a nonlinear Schr{\"o}dinger equation shock example. 
For our purposes here, the matched-asymptotic method is sufficient. 

In this letter: 
We give the required IST results in \S\ref{sec:IST}. 
Then we find the rapidly decaying solution in the region to the right of the DSW (\S\ref{sec:front}). 
This matches into the central DSW (\S\ref{sec:DSW}), which is a slowly varying cnoidal wave with a soliton train on its right and an oscillatory tail on its left. 
Finally, we match the decaying, oscillatory region to the left (\S\ref{sec:tail}) with the DSW. 
We then compare this solution with:
the solution of Burgers' equation (\S\ref{sec:Burgers}), which is the leading-order asymptotic equation for VSWs; 
the solution of the linear KdV equation with step-like data (\S\ref{sec:linearKdV}); 
and the solution of the KdV equation with vanishing data (\S\ref{sec:vanishing}). 
Then we draw some conclusions (\S\ref{sec:conclusion}). 

\section{IST solution} \label{sec:IST}

The IST method transforms the initial data into scattering data, 
evolves the scattering data in time, 
and then recovers the solution from the evolved scattering data. 
First, we associate a linear (Lax) pair with the nonlinear PDE, in this case \eqref{eq:KdV}. 
Then we use the scattering equation of the linear pair to transform the initial data into scattering data. 
The scattering data are then evolved in time using the associated linear equation. 
Finally, the solution is recovered using a linear integral equation, the GLM integral equation, at any time. 

The Lax pair associated with \eqref{eq:KdV} is
\begin{subequations}
\begin{gather} \label{eq:scatter}
	v_{xx} + \left(u/6 +\lambda^2\right) v/\varepsilon^2 = 0, \\
  \label{eq:time}
  v_t = \left(u_x/6 + \gamma\right)v + \left(4\lambda^2 - u/3\right) v_x, 
\end{gather}
\end{subequations}
where $\gamma$ is a constant. 
This linear pair is compatible (that is, $v_{xxt} = v_{txx}$) when $u$ satisfies \eqref{eq:KdV} and $\lambda$ is isospectral (that is, $\partial \lambda/\partial t = 0$). 
The eigenfunctions that satisfy \eqref{eq:scatter} are defined using \eqref{eq:uBC}: 
\begin{subequations} \label{eq:eigenLim}
\begin{equation} \label{eq:phiLim}
	\phi(x;\lambda) \sim e^{-i \lambda x/\varepsilon},\qquad  
	\bar\phi(x;\lambda) \sim e^{i \lambda x/\varepsilon},  
\end{equation}
as $x \to -\infty$ and
\begin{equation} \label{eq:psiLim}
  \psi(x;\lambda_r) \sim e^{i \lambda_r x/\varepsilon}, \quad 
  \bar\psi(x;\lambda_r) \sim e^{-i \lambda_r x/\varepsilon}, 
\end{equation} 
\end{subequations}
as $x \to +\infty$, where $\lambda_r \equiv \sqrt{\lambda^2 - c^2}$.
The branch cut of $\lambda_r$ is taken to be $\lambda\in[-c,c]$;  
the branch cut of $\lambda$ is taken to be $\lambda_r\in[-ic,ic]$: 
so $\im(\lambda_r) \gtrless 0$ when $\im(\lambda) \gtrless 0$. 
This branch cut is one of the main differences between vanishing and step-like data.

The Wronskians $W(\phi,\bar\phi) = 2i\lambda/\varepsilon$ and $W(\psi,\bar\psi) = -2i\lambda_r/\varepsilon$ are constant. 
The scattering eigenfunctions and scattering data $a$ and $b$ associated with \eqref{eq:scatter} satisfy 
\begin{equation} \label{eq:phiDef}
  \phi(x;\lambda) = a(\lambda,\lambda_r) \bar{\psi}(x;\lambda_r) + b(\lambda,\lambda_r) \psi(x;\lambda_r) 
\end{equation} 
for $\lambda_r\neq0$, $\lambda_r \in \mathbb{R}$ (or, equivalently, $|\lambda| > c$, $\lambda\in\mathbb{R}$). 
The scattering data can be written as
$	2i\lambda_r a
	= \varepsilon W(\phi,\psi)$ 
and
$ 2i \lambda_r b = \varepsilon W(\bar\psi,\phi)$. 
We can use this to extend $a$ to $\lambda\in(-c,c)$, where $\lambda_r$ is pure imaginary; 
when $\lambda\in(-c,c)$, $\psi$ is real and exponentially decaying.

We define the transmission coefficient $T \equiv 1/a$ and
the reflection coefficient $R \equiv b/a$ so that \eqref{eq:phiDef} can be written as 
\begin{equation} \label{eq:phiTR}
  T(\lambda,\lambda_r) \phi(x;\lambda) = \bar{\psi}(x;\lambda_r) + R(\lambda,\lambda_r) \psi(x;\lambda_r). 
\end{equation}
From \eqref{eq:uBC}, \eqref{eq:time}, and \eqref{eq:phiTR}, these coefficients evolve in time as 
$ T(\lambda,\lambda_r;t) = T(\lambda,\lambda_r;0)e^{i(4\lambda^2\lambda_r - 4\lambda^3 + 2c^2\lambda_r)t/\varepsilon}$ and 
$R(\lambda,\lambda_r;t) = R(\lambda,\lambda_r;0)e^{i(8\lambda^2\lambda_r+ 4c^2\lambda_r)t/\varepsilon}$.
Unlike with vanishing data, the transmission coefficient, $T$, depends on time; 
this time dependence is not purely phase when $\lambda\in(-c,c)$.

The associated GLM integral equation is
\begin{equation} \label{eq:GLM}
  K(x,y;t) + \Omega(x+y;t) 
    + \int_x^\infty \Omega(y+z;t)K(x,z;t)\,\ud z = 0,
\end{equation}
where 
\begin{multline*}
	\Omega(\xi;t) = \frac{1}{2\varepsilon\pi}\int_{-\infty}^\infty R e^{i\lambda_r \xi/\varepsilon}\,\ud \lambda_r
	+ \sum_j c_j e^{-\tilde \kappa_j \xi/\varepsilon}\\
	+ \frac{1}{2\varepsilon\pi}\int_0^c|\lambda T/\lambda_r|^2 e^{-\sqrt{c^2-\lambda^2}\xi/\varepsilon}\,\ud\lambda, 
\end{multline*}
the constants $\{i\kappa_j\}_{j=1}^N$ are the (simple) poles of $T(i\kappa_j,\lambda_r(i\kappa_j);t)$, 
$\tilde \kappa_j = \sqrt{\kappa_j^2+c^2}$, 
$c_j = -i\mu_j/(\varepsilon \partial_{\lambda_r} a(i\kappa_j))$, 
$\phi(x;i\kappa_j,t) \equiv \mu_j(t) \psi(x;i\kappa_j,t)$, 
and $0<\kappa_1<\dotsb<\kappa_N$ are real. 
We omit any contributions from poles in our asymptotics: 
the poles are related to the solitons, which move to the right, and so do not affect the DSW in the large-time limit. 

From $K$, we recover $u(x,t)$ from
\[	u(x,t) = - 6c^2 + 12 \varepsilon^2 \frac{\ud}{\ud x} K(x,x;t). \]

\section{Long-time asymptotics} \label{sec:asymptotics}

To find the long-time asymptotic solution: 
we find the solution right of the DSW, 
then we use it to find the DSW, and 
then we find the solution left of the DSW and match it into the DSW. 
We use \eqref{eq:GLM} to asymptotically compute the behavior right of the DSW. 
When this asymptotic solution breaks down, we use the matched-asymptotic method introduced in \cite{AblowitzSegur1977} to find the DSW's slowly varying cnoidal solution. 
This naturally leads to Whitham's equations, which were originally obtained by the method of averaging \cite{Whitham1965} and later by a perturbative approach \cite{Luke1966}. 
Then we use the method in \cite{Segur1981} to determine the small, decaying, oscillating solution to the left of the DSW, which matches into the DSW. 

\subsection{Shock front} \label{sec:front}

We asymptotically approximate $u$ to the DSW's right by summing the Neumann series for $K$ in \eqref{eq:GLM}. 
We do this by finding the long-time asymptotic-approximation of $\Omega(\xi;t)$ right of the DSW. 
Then we use the Neumann series formed from the iterates 
$K^{(0)}(x,y;t) = -\Omega(x+y;t)$ and 
$K^{(n)}(x,y;t) = -\Omega(x+y;t) - \int_x^\infty \Omega(y+z;t)K^{(n-1)}(x,z;t)\,\ud z$. 

Far to the DSW's right, the contribution to $\Omega$ from the reflection coefficient dominates and  $[u(x,t) + 6c^2]$ is exponentially small. 

Near the DSW's right, the contribution to $\Omega$ from the transmission coefficient dominates: 
the contribution from $\lambda = 0$ gives 
\[	\Omega \sim 
	\frac{-e^{-ct(\xi/t+4c^2)/\varepsilon}\sqrt{\varepsilon}}%
		{16\sqrt{\pi}[6c - \xi/(2ct)]^{3/2}}\left[
	H_2(0)t^{-3/2}
	+ \mathcal{O}(t^{-5/2}) \right],
\]
where $ H_j(\lambda_*) \equiv [ \partial^j |T(\lambda,\lambda_r(\lambda);0)|^2/\partial \lambda^j ]_{\lambda=\lambda_*}$. 

The terms in the Neumann series become disordered when 
$[ x + 2c^2t + 3\varepsilon/(4c)\log( 6c^2t - x)] = \mathcal{O}(1)$, 
which is at the DSW's right edge 
(cf.\ asymptotic principles discussed in \cite{Kruskal1962}). 
The Neumann series can be summed, and we find that 
\begin{equation} \label{eq:shockFront}
	u(x,t) \sim -6c^2 
		+ 12c^2\sech^2\left[\frac{c}{\varepsilon}(\zeta - \zeta_0)\right],
\end{equation}
where $\zeta_0 = \varepsilon/(2c)\log\{32\pi^{1/2}/[H_2(0)c^{1/2}\varepsilon^{3/2}]\}$ and
\begin{equation}\label{eq:zetaDef}
	\zeta = -x - 2c^2t - \frac{3\varepsilon}{4c}\log (6c^2t-x) + A_1(x/t)t^{-1} + \dotsb. 
\end{equation}
(We omit $A_1$ due to length.)
This provides the boundary condition on the DSW's right. 

This procedure gives the DSW's phase, $\zeta_0$.  
This phase only depends on $H_2(0)$ (since $H_0(0) = H_1(0)=0$); 
the equivalent phase term \cite[Eq.~(2.25c)]{AblowitzSegur1977} for vanishing data is determined by $r''(0) - [r'(0)]^2/r(0)$, where $r$ is the corresponding reflection coefficient. 
The equivalent phase term in the shock solution associated with Burgers' equation also depends on the initial data in a similar way (see \S\ref{sec:Burgers} and \cite{BaldwinPhD}).

\subsection{DSW} \label{sec:DSW}

We find the DSW using matched asymptotics analogous to \cite{AblowitzSegur1977}. 
First we make the variable change $u(x,t) = -6c^2 + g(\zeta,t)$ in \eqref{eq:KdV}, based on \eqref{eq:shockFront}. 
Then we introduce the slow-variables $Z \equiv \delta \zeta$ and $T \equiv \delta t$ (where $\delta = \mathcal{O}(t^{-1})$) to get 
\begin{multline} \label{eq:slowVaryingKdV}
\varepsilon^2 g_{\zeta\zeta\zeta} + gg_\zeta - 4c^2 g_\zeta - g_t \\
	= \delta\left\{
		\frac{3\varepsilon(
			3\varepsilon^2g_{\zeta\zeta\zeta} 
			+ gg_\zeta - 12c^2g_\zeta)}{4c(8c^2T + Z)}
	\right\} + \dotsb. 
\end{multline}
To leading order, \eqref{eq:slowVaryingKdV} has the special solution (which can be found using the methods in \cite{Baldwin2004}) 
\begin{multline} \label{eq:gLocal}
	g(\zeta,t) \sim 4c^2 - V + 4\varepsilon^2\kappa^2(1-2k^2)\\
	 + 12k^2\varepsilon^2\kappa^2 \cn^2\left[\kappa(\zeta-\zeta_0 - Vt),k\right],
\end{multline}
where $\cn(z,k)$ is the  Jacobian elliptic `cosine' (see  \cite{Olver2010}). 
Here, $\kappa$, $k$, and $V$ are arbitrary constants when the right-hand-side of \eqref{eq:slowVaryingKdV} is neglected but vary slowly in general. 
In the special case $k = 1$, $\kappa = c/\varepsilon$, and $V = 0$,  
$ g(\zeta,t) = 12c^2 \sech^2[c(\zeta - \zeta_0)/\varepsilon] $
and exactly matches \eqref{eq:shockFront}. 

As in \cite{Luke1966}, we use the multiple-scales method to determine $\kappa$, $k$, and $V$, which vary with the slow-variables $Z$ and $T$. 
This leads to three conservation laws, which determine $\kappa$, $k$, and $V$. 
First we introduce the rapid-variable $\theta(\zeta,t)$  
with $\theta_\zeta  \equiv \kappa(Z,T)$ and $\theta_t \equiv -\omega(Z,T) \equiv -\kappa V$; 
this leads to the compatibility condition $(\theta_\zeta)_t = (\theta_t)_\zeta$ or 
$ \kappa_T + \omega_Z = 0$ --- which is a conservation law. 
We then use $\partial_t = -\omega\partial_\theta + \delta \partial_T$ 
and $\partial_\zeta = \kappa \partial_\theta + \delta \partial_Z$ 
to transform \eqref{eq:slowVaryingKdV}; 
then we expand $g(\theta,Z,T) = g_0(\theta,Z,T) + \delta g_1(\theta,Z,T) + \delta^2 g_2(\theta,Z,T) + \dotsb$ and group terms in like powers of $\delta$. 
The solution of the $\mathcal{O}(1)$ equation is 
\[	g_0(\theta,Z,T) 
	= a(Z,T) + b(Z,T)\cn^2\{2K(\theta - \theta_0), k(Z,T)\}, \]
where $K\equiv K(k(Z,T))$ is the complete elliptic integral of the first kind,
$ \kappa^2 = b/[48\varepsilon^2k^2K^2]$, and 
$ a = 4c^2 - V - 2b/3 + b/(3k^2)$. 
Then we enforce the periodicity of $g_0(\theta,Z,T)$ in $\theta$ to eliminate secular terms (that is, terms that grow arbitrarily large); 
this gives the other two conservation laws, which we omit due to length. 
These three conservation laws determine $b$, $k$, and $V$. 

If we make the variable change $b/k^2 = 2(r_3-r_1)$, $k^2 = (r_2 - r_1)/(r_3-r_1)$, 
and $V = 4c^2 - (r_1 + r_2 + r_3)/3$, 
then simplifying gives the diagonal system 
\begin{equation} \label{eq:WhithamEqns}
	\frac{\partial r_i}{\partial T} 
	+ v_i(r_1,r_2,r_3)\frac{\partial r_i}{\partial Z} = 0, \qquad i = 1,2,3, 
\end{equation}
where 
$ v_1 = V + bK/[3(K-E)]$,
$ v_2 = V + b(1-k^2)K/[3(E-(1-k^2)K)]$, 
$ v_3 = V - b(1-k^2)K/(3k^2E)$,  
and 
\[	g_0(\theta,Z,T) = r_1 - r_2 + r_3 + 2(r_2 - r_1)\cn^2\left(2K(\theta-\theta_0),k\right). \]
Whitham first found this diagonal system in \cite{Whitham1965} (see also \cite{Gurevich1974,ABH2009}). 

\begin{figure}[htbp]
	\begin{tabular}{cc}
		\includegraphics[width=0.45\hsize]{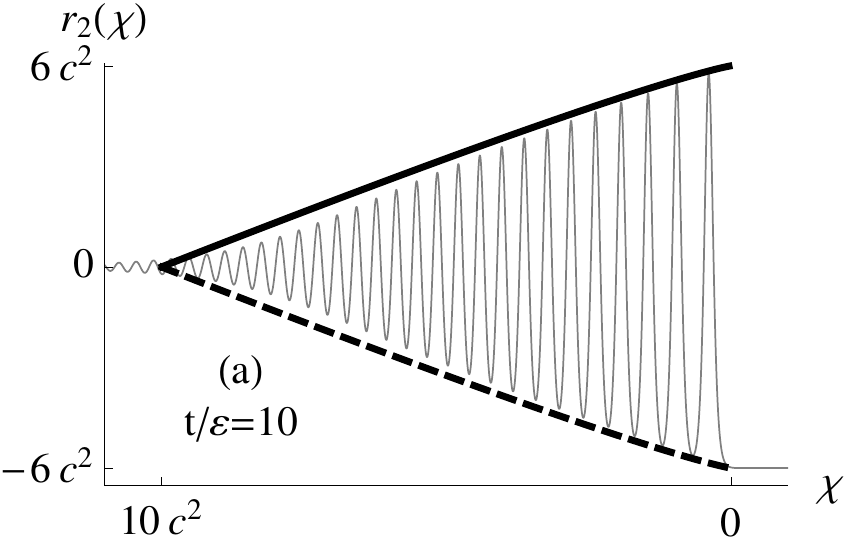} & 
		\includegraphics[width=0.45\hsize]{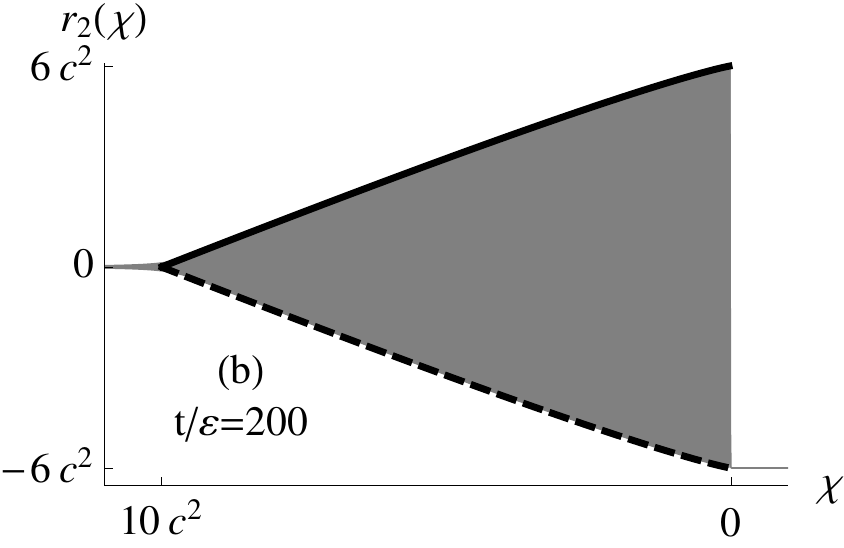} \\
	\end{tabular}
	\caption{\label{fig:r2}The value of $r_2(\chi)$ found numerically for $0 < \chi < 10 c^2$, where $\chi\equiv \zeta/t$. 
   For comparison, we include 
   $-r_2(\chi)$ as a dashed line and 
   a numerical simulation of $u(x,t)$ in gray (inside the envelope of $r_2$ and $-r_2$) for a single-step at (a)~$t/\varepsilon=10$ and (b)~$t/\varepsilon=200$. 
   Note that $\chi = 0$ corresponds to $x \sim -2c^2t$ and $\chi=10c^2$ to $x \sim -12 c^2 t$.}
   \label{fig:self-similar-r2}
\end{figure}

For large-time, the solution tends to a self-similar solution: 
that is, $r_i=r_i(\chi)$ with $\chi \equiv Z/T = \zeta/t$. 
The boundary conditions are satisfied when $r_1 = 0$ and $r_3 = 6c^2$. 
So \eqref{eq:WhithamEqns} reduces to \[ (v_2 - \chi)r_2'(\chi) = 0, \]
which $v_2 = \chi$ satisfies. 
The numerical solution of this implicit equation for $r_2$ is plotted in Fig.~\ref{fig:self-similar-r2}.
We can directly compute the right and left speed of the DSW:
At the DSW's right, we take the limit $r_2 \to r_3$ and get that $v_2 \to 0$ or $x\sim-2c^2t$ --- the speed of the soliton train.
At the DSW's left, we take the limit $r_2 \to r_1$ and get that $v_2 \to 10c^2$ or $x\sim -12c^2t$. 

\subsection{Trailing edge} \label{sec:tail}

On the DSW's left, there is a decaying, slowly varying, oscillatory similarity-solution 
(as there is with vanishing data \cite{AblowitzSegur1977,Segur1981}):
\[	u(x,t) \sim 2A\frac{X^{1/4}}{\sqrt{\tau}} \cos(\theta) 
	- \frac{A^2(1 - \cos 2\theta)}{3\tau\sqrt{X}} 
	+ \mathcal{O}(\tau^{-3/2}), \]
where $X = -x/(3t)$, $\tau = 3t$,
\[	\theta \sim \frac{\tau}{\varepsilon}\left[
		\frac{2}{3}X^{3/2}
		- \frac{A^2}{18}\frac{\log (\tau X^{3/2})}{\tau}
		+ \frac{\theta_0}{\tau}
		+ \mathcal{O}(\tau^{-2})
		\right], \]
and $A$ and $\theta_0$ depend on the scattering data. 

We could, in principle, sum the Neumann series from the GLM integral equation formulated from $-\infty$ to $x$ in the long-time limit to find $A$ and $\theta_0$. 
Indeed, the GLM integral equation formulated from $-\infty$ to $x$ has the same form for both step-like and vanishing data. 

Instead, we use the method in \cite{Segur1981} to find $A$ and $\theta_0$.
Using this oscillatory similarity-solution in \eqref{eq:scatter}, we asymptotically solve for the eigenfunction $\phi$ with boundary-values $\phi\to e^{-i\lambda x/\varepsilon}$ as $x\to-\infty$ and $\phi \to \tilde a e^{-i\lambda x/\varepsilon} + \tilde b e^{i\lambda(x+8\lambda^2 t)/\varepsilon}$ as $x\to-12c^2t$. 
Here, $\tilde a$ and $\tilde b$ can be related to $a$ and $b$ through either the GLM integral equation (formulated from $-\infty$ to $x$) or the asymptotic forms of $u$ for $-12c^2t \ll x \ll \infty$. 
This leads to a matched-asymptotics problem with three regions --- the resonance or turning-point region, where $X \sim 4\lambda^2$, is solved in terms of parabolic cylinder functions. 
The result is that 
\[	A^2(X) \sim 
	- \frac{9\varepsilon}{\pi} \log \left( 1 - \left|R\left(\sqrt{X}/2\right)\right|^2\right) \]
and 
\begin{multline*} 
\frac{\theta_0}{\varepsilon} = \frac{\pi}{4} - \arg\{\tilde r(\lambda,0)\} 
	- \arg\left\{\Gamma\left(1-\frac{iA^2(4\lambda^2)}{18\varepsilon}\right)\right\} \\
	-\frac{c^2A^2(4c^2)}{9\varepsilon\lambda^2} 
			\log\left(\frac{c - \lambda}{c + \lambda}\right)
	-\frac{A^2}{6\varepsilon}\log 2 \\
	- \frac{1}{9\lambda^2\varepsilon}
		\int_c^\lambda \left(\xi^2 A^2(4\xi^2)\right)_\xi
		\log\left(\frac{\xi - \lambda}{\xi + \lambda}\right)\,\ud \xi,
\end{multline*}
where $\lambda = \sqrt{X}/2$, 
$\tilde r(\lambda,t) \equiv \tilde b(\lambda,t)/\tilde a(\lambda)$, 
and $|\tilde r(\lambda,t)| = |R(\lambda,t)|$.
The limits $x \to -12 c^2t$ and $r_2 \sim 2(10c^2-\chi)/3 \sim 2 A X^{1/4} \tau^{-1/2}$ yield $u \sim 2\sqrt{2c/(3t)} \cos(16 c^3 t/\varepsilon)$, which matches the DSW at its left edge (see \cite{BaldwinPhD} for details).

\section{Discussion}

\subsection{Comparison with Burgers' equation} \label{sec:Burgers}

Burgers' equation ($w_t + ww_x - \nu w_{xx}=0$, $\nu>0$) is the leading-order asymptotic equation for VSWs. 
If we take initial data that go rapidly to the boundary conditions $\lim_{x\to-\infty} w(x,t) = 0$ and $\lim_{x\to+\infty} w(x,t) = -h^2$, then the long-time asymptotic solution is 
$ w(x,t) \sim -(h^2/2) \{1 + \tanh[h^2 (x - x_0 + h^2t/2)/(4\nu) ]\}$, 
where $x_0$ is a real constant that depends on the initial data \cite{BaldwinPhD}. 
So well-separated step data go to a single shock wave in the large-time limit for both Burgers' and the KdV equation. 
For both, the boundary data determine its form and the initial data determine its location. 
Unlike with Burgers' equation, the solution of the KdV equation with step-like data can also have a finite number of solitons, which move to the DSW's right in the long-time limit. 

\subsection{Comparison with the linear KdV equation} \label{sec:linearKdV}

The large-time asymptotic solution of \eqref{eq:KdV} differs significantly from the linear problem ($\tilde u_t + \varepsilon^2 \tilde u_{xxx} = 0$). 
While the nonlinear problem has a central DSW region with strong nonlinearity over $|x| = \mathcal{O}(t)$, the linear problem's middle region is only over $|x| = \mathcal{O}(t^{1/3})$. 
Indeed, the solution to the linear problem in the middle region is $\tilde u(x,t) \sim U_0(0)\int_{-\infty}^\eta \Ai(\eta')\,\ud\eta'$, where $\Ai(x)$ is the Airy function, $\eta = x/(3\varepsilon^2t)^{1/3}$, and $U_0$ is the Fourier transform of $\tilde u_x(x,0)$. 

\subsection{Comparison with vanishing boundary conditions} \label{sec:vanishing}

The large-time asymptotic solution of \eqref{eq:KdV} for step-like data ($c\neq 0$) is also significantly different from that for vanishing data ($c=0$). 
The collisionless shock (region~III in Fig.~\ref{fig:decay}b), 
which is analogous to the DSW, has width $\mathcal{O}[t^{1/3}(\log t)^{2/3}]$; 
the DSW (region~B in Fig.~\ref{fig:decay}a) has width $\mathcal{O}(t)$.
From \citet{AblowitzSegur1977}, 
the long-time asymptotic solution for vanishing data have four regions (Fig.~\ref{fig:decay}b) --- all of which decay in time:
an exponentially small solution in region~I, $x \ge \mathcal{O}(t)$;
a growing similarity-solution in region~II, $|x| \le \mathcal{O}(t^{1/3})$, which is related to Painlev\'e II;
a collisionless shock in region~III, $(-x) = \mathcal{O}[t^{1/3}(\log t)^{2/3}]$, which is a slowly varying cnoidal wave analogous to a DSW; 
and an oscillatory similarity-solution in region~IV, $(-x) \ge \mathcal{O}(t)$, which has the same form as the solution in \S\ref{sec:tail}.

\subsection{Conclusion} \label{sec:conclusion}

In this letter, we show that general, step-like initial data tend to a single-phase DSW for large-time. 
Therefore, well-separated, multi-step initial data eventually form a single-phase DSW despite having multiphase dynamics at intermediate times (as indicated in Fig.~\ref{fig:numerics}).
The asymptotic solution of the KdV equation for general, step-like data is new. 
The details of our calculations will be given in a separate paper; 
we anticipate that they can be applied to other integrable nonlinear PDEs with general, step-like data.

\section*{Acknowledgements}
We wish to acknowledge the support of the National Science Foundation under grant DMS-0905779 and the Air Force Office of Scientific Research under grant FA9550-12-1-0207.

\bibliographystyle{model1a-num-names}

\end{document}